\documentclass[runningheads]{llncs}
\usepackage[T1]{fontenc}

\usepackage{svg}
\usepackage{cite}
\usepackage{booktabs}
\usepackage{hyperref}
\usepackage{graphicx}
\usepackage{comment}
\usepackage{threeparttable}
\usepackage{color}

\hypersetup{
    pdftitle={Large Kernel MedNeXt for Breast Tumor Segmentation and Self-Normalizing Network for pCR Classification in Magnetic Resonance Images},
    pdfauthor={Toufiq Musah},
    pdfkeywords={Breast Cancer, Segmentation, Radiomics, Deep Learning, MRI, pCR, MedNeXt, Self-Normalizing Network}
}

\begin{document}

\title{Large Kernel MedNeXt for Breast Tumor Segmentation and Self-Normalizing Network for pCR Classification in Magnetic Resonance Images}

\titlerunning{Large Kernel MedNeXt for Breast Tumor Segmentation}

\authorrunning{T. Musah}

\author{Toufiq Musah\inst{1, 2}}

\institute{
Department of Computer Engineering, Kwame Nkrumah University of Science and Technology, Kumasi, Ghana.\\ \and
Global Health and Infectious Disease Group, Kumasi Centre for Collaborative Research in Tropical Medicine, Kumasi, Ghana. \\
\email{toufiqmusah32@gmail.com}}

\maketitle              

\begin{abstract}
Accurate breast tumor segmentation in dynamic contrast-enhanced magnetic resonance imaging (DCE-MRI) is important for downstream tasks such as pathological complete response (pCR) assessment. In this work, we address both segmentation and pCR classification using the large-scale MAMA-MIA DCE-MRI dataset. We employ a large-kernel MedNeXt architecture with a two-stage training strategy that expands the receptive field from $3 \times 3 \times 3$ to $5 \times 5 \times 5$ kernels using the UpKern algorithm. This approach allows stable transfer of learned features to larger kernels, improving segmentation performance on the unseen validation set. An ensemble of large-kernel models achieved a Dice score of 0.67 and a normalized Hausdorff Distance (NormHD) of 0.24. For pCR classification, we trained a self-normalizing network (SNN) on radiomic features extracted from the predicted segmentations and first post-contrast DCE-MRI, reaching an average balanced accuracy of 57\%, and up to 75\% in some subgroups. Our findings highlight the benefits of combining larger receptive fields and radiomics-driven classification while motivating future work on advanced ensembling and the integration of clinical variables to further improve performance and generalization. \textbf{Code}: \url{https://github.com/toufiqmusah/caladan-mama-mia.git}

\end{abstract}
\section{Introduction}
Breast cancer is a major health concern \cite{burden}, and Magnetic Resonance Imaging (MRI) plays an important role in its assessment, especially for preoperative staging monitoring treatment response, and pathological complete response (pCR) assessment \cite{role, pcr-mri-1, pcr-mri-2}. T1-weighted dynamic contrast-enhanced MRI (DCE-MRI) can highlight tumor vascularity by using contrast agents, aiding in the localization of malignant lesions. Accurate segmentation of tumor boundaries on these images is clinically valuable, as it enables quantitative evaluation of tumor size, shape, and volume over time \cite{prognostic, gaps}. High-quality tumor segmentation further facilitates advanced analyses such radiomics feature extraction \cite{radiomics-OG} for other downstream tasks \cite{rad-1, rad-2}, including pCR assessment \cite{pcr-rad}. There is however difficulty in obtaining expert-labeled tumor masks for breast MRI \cite{lidia}. 

Recently, a large-scale multi-center breast DCE-MRI dataset was released to address the data scarcity. Garrucho et al. \cite{lidia} introduced the MAMA-MIA challenge and dataset, the largest of its kind, comprising 1506 pre-treatment DCE-MRI cases with expert segmentations of primary breast tumors and non-mass enhancement, along with class labels for pCR status.

Deep learning architectures such as U-Net \cite{unet} and its variants \cite{milletari2016v, cciccek20163d, oktay2018attention} have been particularly successful in biomedical image segmentation. These models typically adopt conventional CNN design principles, using $3 \times 3$ convolutional kernels to balance computational efficiency and parameter count \cite{scaling-kernels}. While effective, small kernels constrain the receptive field, potentially limiting the model’s ability to capture long-range context, which is an important consideration in whole-breast MRI volumes \cite{mednext}. To address this, recent research has explored Transformer-based architectures in medical imaging \cite{vit}. Vision Transformers and hybrid Transformer-CNN models have achieved state-of-the-art performance across various segmentation tasks by modeling long-range dependencies explicitly \cite{swin}. However, Transformers generally require large-scale training datasets to generalize well, as they lack the strong spatial inductive biases inherent to CNNs \cite{vit-understand}. Further, the task of classifying pCR has proven difficult in both imaging and non-imaging approaches \cite{difficult-pcr, difficult-pcr-2}. Radiomics \cite{radiomics-OG} has proven to be capable in extracting semantic features for such downstream tasks \cite{pcr-rad, rad-1}, though feature distributions are highly abstract. Self-Normalizing Networks (SNNs) \cite{SNN} were introduced to mitigate this problem by having neurons that automatically converge to zero mean and unit variance without the explicit need for normalization layers.

In this work, we propose a large-kernel MedNeXt architecture \cite{mednext} for breast tumor segmentation, and a radiomics-based SNN for pathological complete response (pCR) classification in DCE-MRI. Our segmentation method is motivated by the need to capture both the subtle enhancement patterns of tumors across multiple post-contrast time points and the broader breast tissue context, which larger receptive fields naturally accommodate. To effectively utilize large kernels without overfitting, we adopt a two-stage training strategy: we first train a MedNeXt model with conventional $3 \times 3 \times 3$ kernels and then expand to $5 \times 5 \times 5$ kernels using the UpKern algorithm \cite{mednext}. UpKern initializes large-kernel networks by trilinear interpolation of the smaller-kernel model’s convolutional filters, mitigating the performance degradation observed when large kernels are trained from scratch. For pCR prediction, we extract radiomic features from the predicted segmentations and DCE-MRI data and train an SNN classifier to stabilize feature learning without explicit normalization layers. This combined approach targets both robust tumor segmentation and reliable downstream pCR assessment.

% Lit sectionIn the realm of breast MRI segmentation, some studies have also attempted to utilize the temporal dimension of DCE-MRI: early approaches simplified the 4D (3D + time) data by using 2D slices or maximum-intensity projections 14 , whereas more recent methods aim to process the full temporal sequence with specialized modules 15 16. For instance, Zhou et al. (2022) proposed a multi-branch 3D network to learn affinities across time points for DCE-MRI tumor segmentation 16 , and Zhao et al. (2024) introduced a Temporal-Spatial Enhanced Network (TSESNet) that fuses features from multiple time points to capture temporal context, achieving leading accuracy on smaller DCE-MRI datasets 17 18 . These innovations underscore the importance of leveraging both spatial and temporal information for improved segmentation of breast tumors.

%

\section{Materials and Methods}

% Maybe mention how large kernel sizes is depth-wise conv & upkern approximate attention ..

\subsection{Dataset}
\begin{figure}
    \centering
    \includegraphics[width=1.0\textwidth]{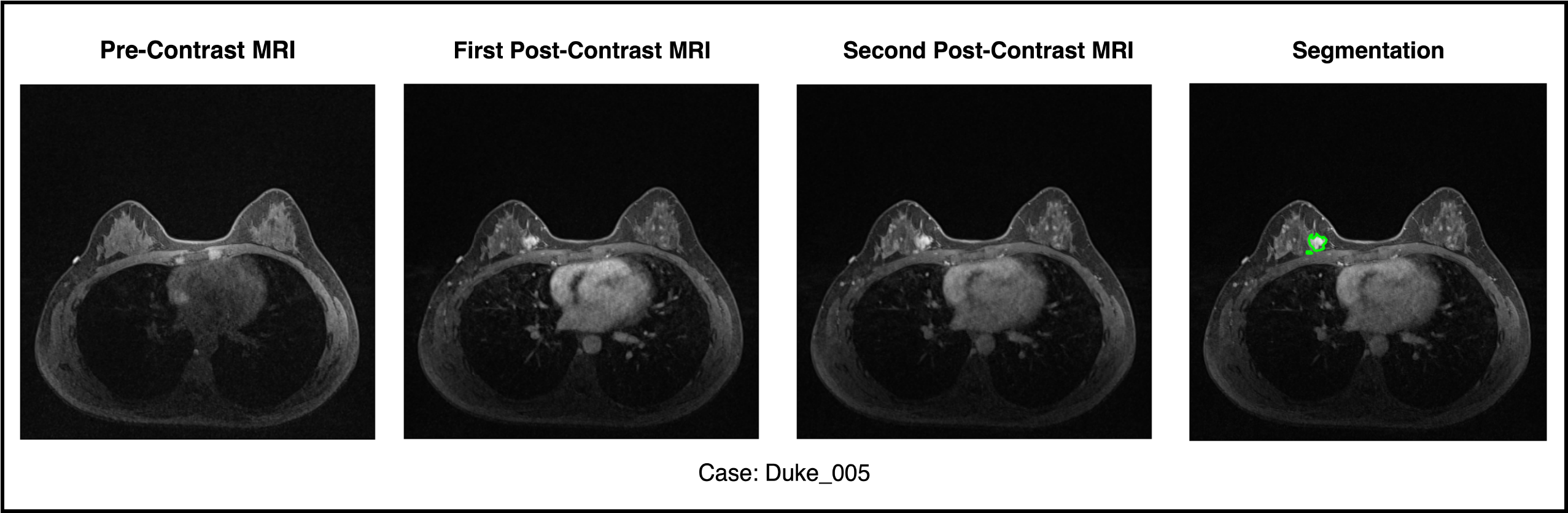}
    \caption{Sample DCE-MRI data showing pre-contrast and post-contrast scans along with expert segmentation in green}
    \label{fig:dataset}
\end{figure}

We used the MAMA-MIA DCE-MRI dataset \cite{lidia} of 1506 cases for training. Each case in the dataset includes a series of T1-weighted DCE-MRI volumes acquired at multiple time points: one pre-contrast and up to 5 post-contrast phases. For our experiments, we selected the pre-contrast image and the first two post-contrast images for each case as seen in Figure \ref{fig:dataset}. Preprocessing and data handling were performed using the nnUNet framework \cite{isensee2021nnu}. The images were resampled to an isotropic voxel spacing of 1.0~mm in all axes, in line with the dataset’s original spacing. Input patches of size $128 \times 128 \times 128$ voxels were used for model training. 

\subsection{MedNeXt Segmentation Training}

\begin{figure}
    \centering
    \includegraphics[width=1.0\textwidth]{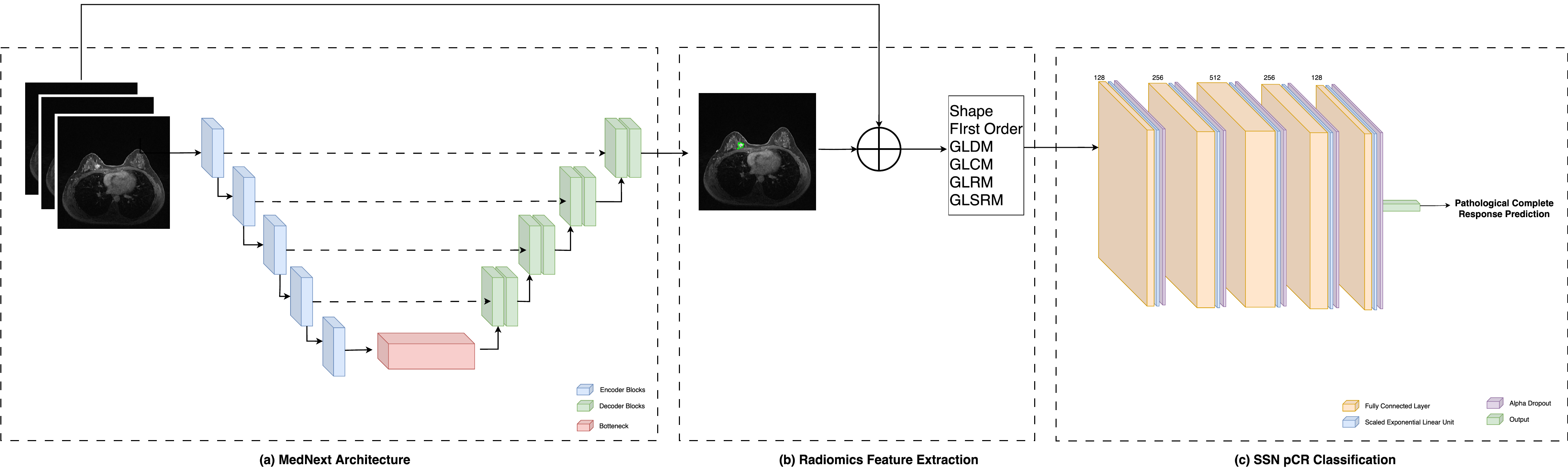}
    \caption{Overview of the proposed pipeline. (a) MedNeXt performs tumor segmentation on DCE-MRI. (b) Radiomic features are extracted from the segmentations and first post-contrast MRI. (c) A Self-Normalizing Network (SNN) predicts pathological complete response (pCR) from the selected features.}
    \label{fig:methods}
\end{figure}

We employed the medium variant of MedNeXt, providing a balance between model capacity and computational efficiency for a dataset of this scale. Our modeling strategy consisted of a two-stage process. In the first stage, we trained the MedNeXt model configured with $3 \times 3 \times 3$ convolutional kernels ($M^3$) with deep supervision, stochastic gradient descent optimization, a cosine annealing learning rate, and Dice-cross-entropy loss. The training was conducted using a 5-fold cross-validation split. Each fold was trained independently for 250 epochs. Following the completion of cross-validation, we identified the fold that achieved the highest mean Dice coefficient on its respective validation set, whose weights were used in the second stage of training.

In the subsequent stage, we employed the UpKern strategy to resample the learned $3 \times 3 \times 3$ kernel weights into $5 \times 5 \times 5$ kernels ($M{^5}{_{Base}}$) via trilinear interpolation. This approach enabled a smooth transition to a large-kernel configuration, thus expanding the effective receptive field of the network without introducing instability often associated with training large kernels from random initialization. The newly initialized large-kernel MedNeXt-M model was then fine-tuned for making use of the entire training set, with all other architectural and training settings held constant.

Additionally, we generated a second large-kernel model ($M{^5}{_{Focal}}$) by repeating the UpKern procedure and fine-tuning, this time using a combination of Dice-cross-entropy and Focal loss to increase penalization of small lesion segmentation errors.

\subsection{pCR Classification}

For pCR classification, we extracted radiomic features from tumor segmentations and the first post-contrast MRI (Figure \ref{fig:methods} (b)). To reduce redundancy, we first removed highly correlated features based on pairwise correlation analysis(threshold >90\%). A sequential feature selector was then employed to identify an informative subset, resulting in 40 features spanning shape, first-order statistics, and texture descriptors.

The selected features included shape metrics such as surface area, elongation, and axis lengths; first-order statistics capturing intensity distributions, including entropy, kurtosis, and total energy; and texture-based descriptors derived from gray-level co-occurrence matrix (GLCM), gray-level size zone matrix (GLSZM), gray-level run length matrix (GLRLM), gray-level dependence matrix (GLDM), and neighboring gray tone difference matrix (NGTDM).

We trained a Self-Normalizing Network (SNN) classifier \cite{SNN} on the selected radiomic features (Figure \ref{fig:methods} (c)). The SNN architecture was chosen for its ability to maintain stable activations without explicit normalization layers, which is particularly advantageous when working with radiomics data that may present diverse value ranges and distributions. The model was trained using binary cross-entropy loss to predict pCR status.

\section{Results}

\subsection{Segmentation Performance}
The in-training validation results for $M^3$ across the 5-fold cross-validation are presented in Table \ref{tab:in_training_dice}.

\begin{table}[!htbp]
    \centering
    \caption{In-Training Validation Dice Scores for $M^3$}
    \label{tab:in_training_dice}
    \begin{tabular}{lc}
        \toprule
        \textbf{Fold} & \textbf{Dice Score} \\
        \midrule
        Fold 0 & 0.786 \\
        Fold 1 & 0.750 \\
        Fold 2 & 0.828 \\
        Fold 3 & 0.795 \\
        Fold 4 & 0.790 \\
        \midrule
        Average & 0.789 \\
        \bottomrule
    \end{tabular}
\end{table}

\noindent Performance of $M^3$ was further evaluated on the unseen validation set on the Codabench platform \footnote{\url{https://www.codabench.org/competitions/7425/\#/results-tab}}. The results are summarized in Table \ref{tab:codabench_results}. $M^3$ achieved a Dice score of 0.64 and a normalized Hausdorff Distance (NormHD) of 0.3. Upon applying the UpKern strategy and fine-tuning the best performing single model, $M{^5}{_{Base}}$ improved the Dice score to 0.66 and the NormHD to 0.29. Further ensembling with $M{^5}{_{Focal}}$ led to minimal increase in Dice to 0.67, and NormHD to 0.24

\begin{table}[!htbp]
    \centering
    \caption{Performance comparison between models. Metrics are reported as Dice Score and Normalized Hausdorff Distance}
    \label{tab:codabench_results}
    \begin{tabular}{l@{\hspace{1em}}c@{\hspace{1em}}c}        \toprule
        \textbf{Model Configuration} & \textbf{Dice Score ($\uparrow$)} & \textbf{NormHD ($\downarrow$)} \\
        \midrule
        $M^3$, (5-Folds) & 0.64 & 0.30 \\
        $M{^5}{_{Base}}$ & 0.66 & 0.29 \\
        $M{^5}{_{Base}}$ + $M{^5}{_{Focal}}$ & \textbf{0.67} & \textbf{0.24} \\
        \bottomrule
    \end{tabular}
\end{table}

\subsection{pCR Classification Performance}
The Self-Normalizing Network (SNN) classifier achieved a balanced accuracy of 57\% on the unseen validation set. The model outputs were highly confident, with most predictions clustered at 0 or 1, suggesting potential overconfidence \cite{over}. Subgroup analysis revealed variations in performance across demographic and clinical variables. For example, balanced accuracy ranged from 75\% in the 51–60 age group to 30\% in the 71+ age group, and from 75\% in breast density group 'a' to 41.7\% in density group 'd', indicating performance disparity across subpopulations.

\section{Discussion}
In-training validation Dice score of $M^3$ averaged \textbf{0.7899 $\pm$ 0.012} across 5 folds, indicates stable learning and internal consistency in cross-validation.

On the unseen validation set, the $M^3$ 5-Fold ensemble achieved a Dice score of 0.64 and a NormHD of 0.30. While this represents a reasonable baseline, the performance gap relative to internal validation suggests the presence of increased data variability in the external set. Performance improved after applying the UpKern strategy to expand the receptive field. Fine-tuning the best-performing fold into $M{^5}_{{Base}}$ led to a Dice score of 0.66 and a reduction in NormHD to 0.29. Further ensembling $M{^5}_{{Base}}$ with $M{^5}_{{Focal}}$, which was trained using Dice-cross-entropy and Focal loss, resulted in a Dice score of 0.67 and a NormHD of 0.24. These results validate the hypothesis that larger receptive fields improve segmentation performance by capturing broader anatomical context.

For pathological complete response (pCR) classification, the self-normalizing network trained on radiomic features achieved a balanced accuracy of 57\% on the unseen validation set. Output inspection indicates a tendency toward overconfident binary predictions with minimal probabilistic spread, which points to a calibration issue. Post-hoc temperature scaling is currently being explored as a corrective measure to soften decision boundaries and produce more reliable probability estimates.

Subgroup analysis revealed performance disparities across fairness variables. Balanced accuracy ranged from 75\% in the 51-60 age group to 30\% in patients over 71, and from 75\% in breast density group `a' to 41.7\% in density group `d'. These gaps highlight the need for future work focused on improving model generalization across diverse demographic and physiological subgroups, potentially through the integration of radiomic features with relevant clinical variables such as age, menopausal status, and breast density.

Additional efforts will focus on integrating calibration techniques and fairness-aware optimization to improve robustness and reduce subgroup performance disparities in both segmentation and pCR classification.

\section{Conclusion}
This work presented a large-kernel MedNeXt architecture, combined with the UpKern strategy, for breast tumor segmentation on the MAMA-MIA DCE-MRI dataset. Expanding the kernels from $3 \times 3 \times 3$ to $5 \times 5 \times 5$ improved segmentation performance on the unseen validation set, confirming the benefit of larger receptive fields.

For pCR classification, a self-normalizing network trained on radiomic features achieved a balanced accuracy of 57\%. The results highlight the need for better calibration and generalization, particularly across demographic subgroups. Future work will explore advanced ensembling and the integration of clinical variables such as age, menopausal status, and breast density to address these challenges.

\bibliographystyle{splncs04}
\bibliography{references}

\end{document}